\documentclass[runningheads]{llncs}
\usepackage{graphicx}
\usepackage{todonotes}
\usepackage{multirow}
\usepackage{cite}
\usepackage{amsmath,amssymb}
\usepackage{bbm}
\usepackage{dsfont}
\usepackage{mathtools}
\usepackage{adjustbox}

\newcommand{\func}[1]{\uppercase{\textit{#1}}}
\begin{document}
\title{Disentangled Representations for Domain-generalized Cardiac Segmentation}

\titlerunning{Disentangled Representations for Domain-generalized Cardiac Segmentation}

\author{Xiao Liu\inst{1}, Spyridon Thermos\inst{1}, Agisilaos Chartsias\inst{1}, Alison O'Neil\inst{1,3} \and
Sotirios A. Tsaftaris\inst{1,2}} 
\authorrunning{X. Liu et al.}

\institute{School of Engineering, University of Edinburgh, Edinburgh EH9 3FB, UK \and
The Alan Turing Institute, London, UK \and
Canon Medical Research Europe Ltd., Edinburgh, UK\\
\email{Xiao.Liu@ed.ac.uk, SThermos@ed.ac.uk, Agis.Chartsias@ed.ac.uk, Alison.ONeil@eu.medical.canon, S.Tsaftaris@ed.ac.uk}}
\maketitle              
\begin{abstract}
Robust cardiac image segmentation is still an open challenge due to the inability of the existing methods to achieve satisfactory performance on unseen data of different domains. Since the acquisition and annotation of medical data are costly and time-consuming, recent work focuses on domain adaptation and generalization to bridge the gap between data from different populations and scanners. In this paper, we propose two data augmentation methods that focus on improving the domain adaptation and generalization abilities of state-to-the-art cardiac segmentation models. In particular, our ``Resolution Augmentation" method generates more diverse data by rescaling images to different resolutions within a range spanning different scanner protocols. Subsequently, our ``Factor-based Augmentation" method generates more diverse data by projecting the original samples onto disentangled latent spaces, and combining the learned anatomy and modality factors from different domains. Our extensive experiments demonstrate the importance of efficient adaptation between seen and unseen domains, as well as model generalization ability, to robust cardiac image segmentation.\footnote{The code will be made publicly available.}

\end{abstract}

\section{Introduction}
\label{Sec::Introduction}
Recent advances in machine and deep learning, as well as in computational hardware, have significantly impacted the field of medical image analysis \cite{litjens2017survey}, and more specifically the task of cardiac image segmentation \cite{chen2020frontiers}. Since cardiovascular disease continues to be a major public health concern, accurate automatic segmentation of cardiac structures is an important step towards improved analysis for prevention and diagnosis. Most existing deep learning methods rely on fully or semi-supervised setups, using pixel-level annotations provided by clinical experts for model training. These methods achieve satisfactory performance in predicting left ventricle (LV), right ventricle (RV), and left ventricular myocardium (MYO) binary masks from unseen data within the same domain. However, such models struggle to achieve similar performance on unseen data from other domains, \textit{e.g.}~captured by another scanner or from a different population \cite{tao2019deep}.

Better generalization of models is an open problem in computer vision \cite{ghifary2015domain, ganin2015unsupervised}. Learning a model with good generalization to the target domain is termed \emph{domain adaptation} when labeled samples from the source domain and unlabeled samples from the target domain are given, or \emph{domain generalization} when only labeled samples from the source domain are given.

\begin{figure}[t]
\includegraphics[width=\textwidth]{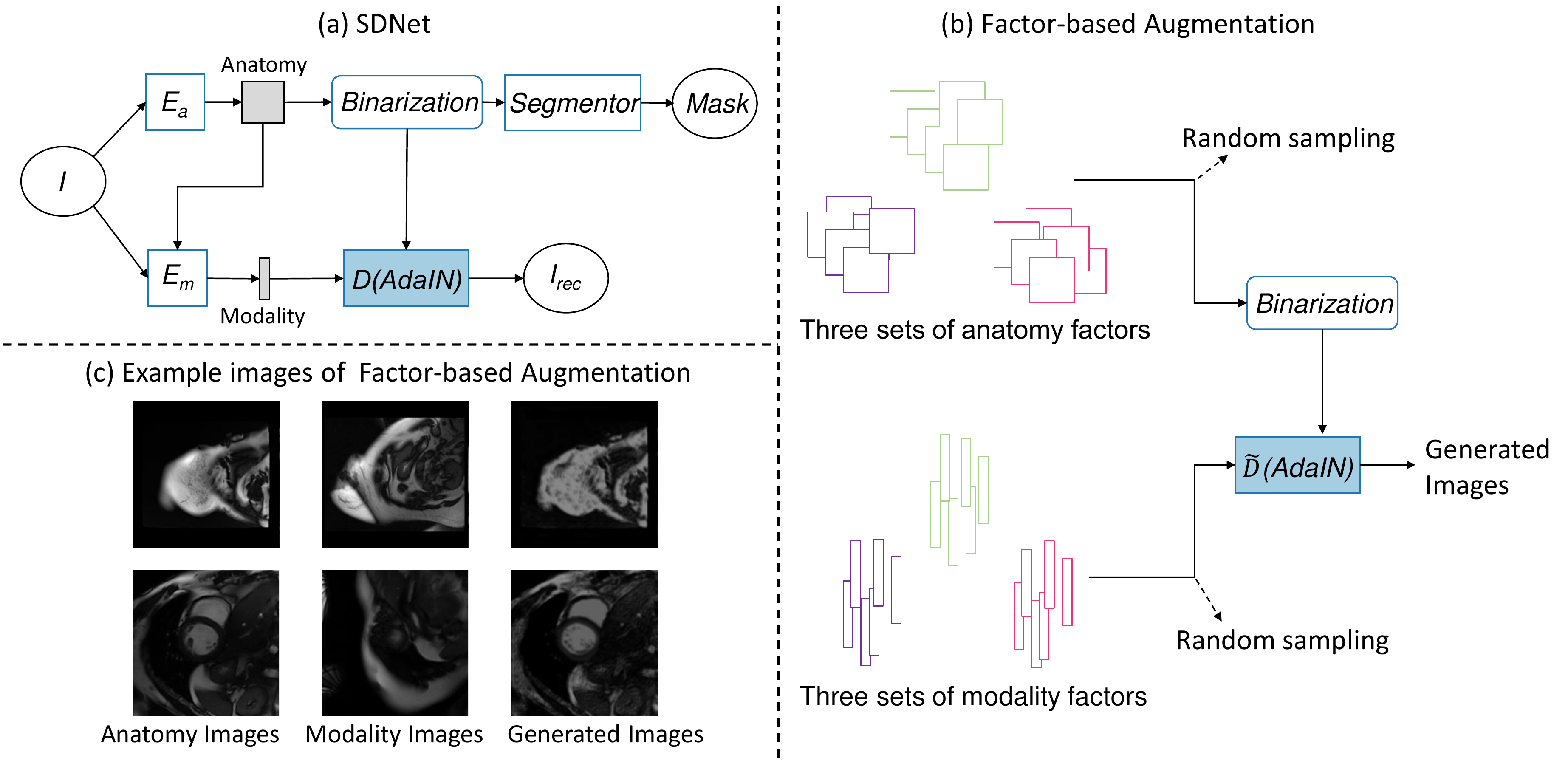}
\vspace{-0.1in}
\caption{(a) SDNet: $\func{E}_a$: anatomy encoder, $\func{E}_m$: modality encoder, $\func{D}(AdaIN)$: AdaIN decoder. $I$ is the input image to the model and $I_{rec}$ is the output image of the AdaIN decoder \textit{i.e.}~the reconstructed image. $Mask$ is the predicted segmentation mask for the input image. (b) Illustration of Factor-based Augmentation: $\tilde{D}(AdaIN)$ is a pre-trained AdaIN decoder. (c) Example images produced by Factor-based Augmentation. Anatomy Images provide anatomy factors,  Modality Images provide modality factors, and Generated Images are the combination of the anatomy and modality factors.} \label{fig:1:overview}
\vspace{-0.1in}
\end{figure}

Annotating more medical images is difficult due to time, cost, and privacy constraints. Therefore, researchers focus on bridging the gap between different domains by learning more general representations from the available data in a semi-supervised manner, \textit{i.e.}~domain adaptation \cite{jiang2018tumor, yang2019domain}. Existing methods in cardiac image segmentation achieve domain generalization either by disentangling spatial (anatomy) from imaging (modality) factors~\cite{sdnet}, or by augmenting the available data \cite{chaitanya2019ipmi, zhao2019cvpr, chen2020miccai}. However, there is limited work~\cite{hsu2019icassp} focusing on mixing disentangled representations, \textit{i.e.}~factors, to augment the original data. 

In this paper, we propose two data augmentation methods, termed Resolution Augmentation (RA) and Factor-based Augmentation (FA), which are combined to improve domain adaption and generalization, thus boosting the performance of state-of-the-art models in Cardiac Magnetic Resonance (CMR) image segmentation. In particular, we use RA to remove the resolution bias by randomly rescaling the training images within a predetermined resolution range, while FA is used to increase diversity in the labeled data through mixing spatial and imaging factors, which we denote as the \textit{anatomy} and \textit{modality} factors, respectively. To extract these factors for FA, we pre-train the SDNet model introduced in~\cite{sdnet} using the original (prior to augmentation) data. Experiments on the diverse dataset from the STACOM 2020 Multi-Centre, Multi-Vendor $\&$ Multi-Disease Cardiac Image Segmentation Challenge (M$\&$Ms challenge) show the superiority of the proposed augmentation methods when combined with the state-of-the-art U-Net~\cite{ronneberger2015u} and SDNet models.

\section{Method}
\label{Sec::Method}
We train the SDNet model and employ RA and FA to generate a more diverse dataset. The model and FA setup is illustrated in Fig.~\ref{fig:1:overview}.

\begin{figure}[t]
\includegraphics[width=\textwidth]{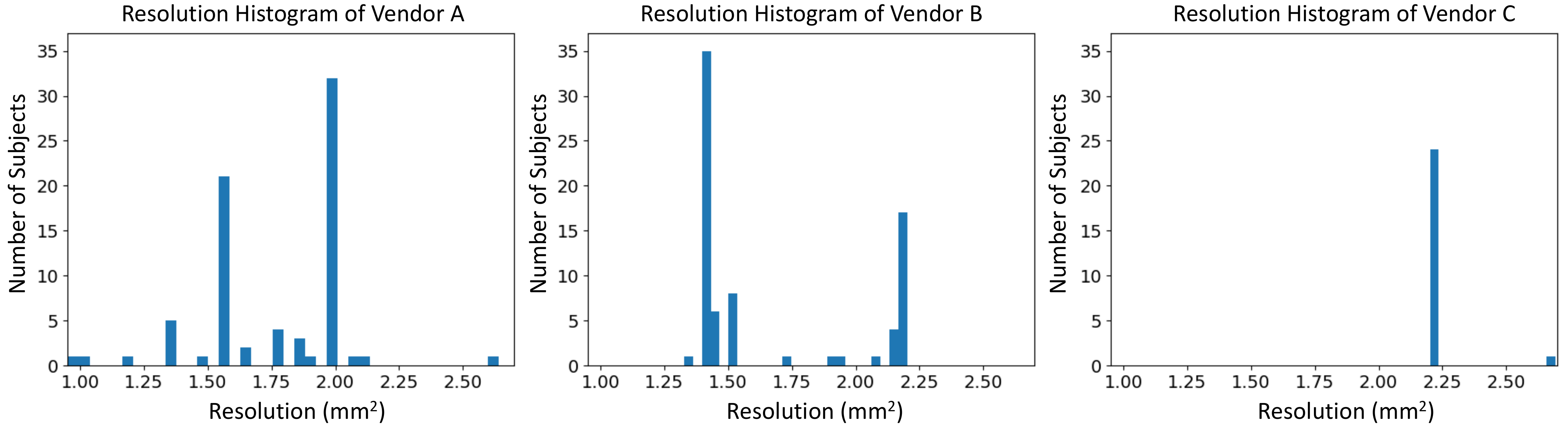}
\vspace{-0.2in}
\centering
\caption{Resolution histograms of the M$\&$Ms challenge training data, broken down by vendor (from left to right: Vendors A, B and C).} \label{fig:2:resolutions}
\end{figure}

\subsection{Proposed augmentation methods}

\subsubsection{Resolution Augmentation (RA):} It is common in MRI data that the imaging resolution is different for each study due to variation of the scanner parameters. Variation in the imaging resolution can cause the cardiac anatomy to vary significantly in size (\textit{i.e.}~area in pixels), beyond normal anatomical variability.

The training dataset contains subjects scanned by scanners from three vendors \textit{i.e.}~Vendors A, B and C. In Fig. \ref{fig:2:resolutions}, we show histograms of the training dataset image resolutions (Sec. \ref{Sec::Experiments} has more details on the dataset). We observed that the histograms of subjects imaged by scanners from different vendors are distinct from one other \textit{i.e.}~this is a bias with respect to the dataset. To reduce this bias, we propose to augment the training dataset such that the resolutions of subjects are equally distributed from 0.954 mm$^2$ to 2.692 mm$^2$ per pixel (the minimum and maximum values observed in the data from the 3 vendors), by rescaling the original image to a random resolution in this range. Finally we center-crop the rescaled image to uniform dimensions of $224 \times 224$ pixels. 

\subsubsection{Factor-based Augmentation (FA):} As shown in Fig.~\ref{fig:1:overview}(a), SDNet decomposes the input image into two latent representations, \textit{i.e.}~anatomy and modality factors, respectively. In particular, the anatomy factor contains spatial information about the input, and the modality factor contains non-spatial information only, namely imaging specific characteristics. Ideally, the anatomy factors would not encode the variation caused by different scanners, rather this information would be encoded in the modality factors. Motivated by this, we propose to augment the training dataset by combining different anatomy and modality factors to generate new data.

Considering the three sets of data from Vendors A, B and C, we first pre-train a SDNet model in a semi-supervised manner using the original data. Using this model at inference, we decompose the three sets of data into three sets of anatomy and modality factors. As shown in Fig. \ref{fig:1:overview}(b), we randomly sample an anatomy factor from the three anatomy factor sets and a modality factor from the three modality factor sets. A new image can be generated by processing the two factors with the decoder of the pre-trained SDNet model. By repeating this augmentation process, we generate a larger and more diverse dataset that contains segmentation annotations. The segmentation mask of the generated data is the mask of the image providing the anatomy factor, \emph{if} the image is labeled, otherwise the generated data is unlabeled. Some indicative examples of FA are visualized in Fig.~\ref{fig:1:overview}(c).

\subsection{Model Architecture}
As depicted in Fig.~\ref{fig:1:overview}(a), the SDNet model consists of 4 basic modules, namely the anatomy encoder $\func{E}_a$, the modality encoder $\func{E}_m$, the segmentor, and the AdaIN-based decoder $\func{D}(AdaIN)$.

The \textbf{anatomy encoder} is realized as a U-Net network that consists of 4 downsampling and upsampling convolutional layers coupled with batch normalization~\cite{ioffe2015batch} layers and ReLU~\cite{nair2010rectified} non-linearities. The output feature of the anatomy encoder has 8 channels, while the feature values are thresholded to 0 and 1 by a differentiable rounding operator. By adopting the thresholding constraint and supervision provided by the segmentation masks, the encoded anatomy factor is forced to contain more spatial information about the image.

The \textbf{modality encoder} consists of 2 downsampling convolutional layers ($4\times4$ kernel size) that are followed by a global averaging pooling layer, which is used to eliminate the spatial (anatomical) information. The output of the pooling layer is then projected to an 8-dimensional vector (modality factor) using a Multi-Layer Perceptron (MLP) network.

The \textbf{segmentor} has 2 convolutional layers using $3\times3$ kernels, coupled with batch normalization and ReLU activation layers, as well as a $1\times1$ convolution followed by a channel-wise softmax function. The input to the segmentor is the thresholded anatomy factor. The output target of this segmentor is the ground-truth segmentation masks when the masks are available. In the learning process, this segmentor encourages the anatomy encoder to encode more spatial information about the image such that the segmentor can learn to predict the corresponding masks more efficiently.

Finally, for the \textbf{AdaIN decoder}, we use the AdaIN module as in~\cite{huang2017arbitrary}, in order to combine the anatomy and modality representations to reconstruct the image. In particular, the decoder consists of 3 convolutional layers ($3\times3$ kernel size) coupled with adaptive instance normalization and ReLU activation layers. A final convolutional layer with $7\times7$ kernels is used for the reconstruction, followed by a hyperbolic tangent activation function that normalizes the values of the generated image into the [0,1] range. As discussed in~\cite{huang2017arbitrary}, the AdaIN decoder normalizes the anatomy factor by firstly applying instance normalization to significantly remove the non-spatial information, then allowing the modality factor to define the new mean and standard derivation of the normalized anatomy factor. In this way, the decoder encourages the anatomy factor to contain spatial information and also force the modality factor to contain non-spatial information. On the other hand, by using AdaIN, the decoder does not simply ignore the modality factor that has a much smaller dimensionality than that of the anatomy factor.

\subsection{Objective Function and Model Training}
Apart from the original SDNet objective functions, we additionally use focal loss as presented in~\cite{lin2017focal}. Focal loss is widely used in segmentation tasks and helps the model to achieve better accuracy by addressing the class imbalance problem. The augmented overall objective function is defined as:

\begin{equation}
    \mathcal{L}_{total} = \lambda_{1} \mathcal{L}_{rec} + \lambda_{2} \mathcal{L}_{z_{rec}} + \lambda_{3} \mathcal{L}_{dice} + \lambda_{4} \mathcal{L}_{focal},
\end{equation}
where $\mathcal{L}_{rec}$ is the $L1$ distance between the input and the reconstructed image, $\mathcal{L}_{z_{rec}}$ denotes the $L1$ distance between the sampled and the re-encoded modality vector (latent regression), $\mathcal{L}_{dice}$ is the segmentation dice loss~\cite{sudre2017generalised}, and $\mathcal{L}_{focal}$ is the segmentation focal loss. Since we train SDNet in both fully supervised  and semi-supervised setups, we set the hyperparameters $\lambda_{1}=\lambda_{2}=\lambda_{3}=\lambda_{4}=1$ when training using labeled data, while when training using unlabeled data we set $\lambda_{1}=\lambda_{2}=1$ and $\lambda_{3}=\lambda_{4}=0$.

\section{Experiment}
\label{Sec::Experiments}
\sloppy
\subsection{Dataset description and preprocessing}
We train and validate our proposed method on the M$\&$Ms challenge dataset of 350 subjects. Some subjects have hypertrophic and dilated cardiomyopathies (and some are healthy) but disease labels are not provided. Subjects were scanned in clinical centres in 3 different countries using 4 different magnetic resonance scanner vendors \textit{i.e.}~Vendor A, B, C and D in this paper. The M$\&$Ms Challenge training dataset contains 75 \textbf{labeled} subjects scanned using technology from Vendor A, 75 \textbf{labeled} subjects scanned by Vendor B, and 25 \textbf{unlabeled} subjects scanned by Vendor C.\footnote{We denote subjects scanned by scanners of Vendor A and B as labeled. However, it is notable that only the end-diastole and end-systole phases are labeled.}
The M$\&$Ms challenge test dataset contains 200 subjects (50 from each vendor, including the seen 25 unlabeled subjects from Vendor C). From these, 80 subjects (20 from each vendor) are used to validate the model (results reported in Table \ref{tab1}) and the rest will be used for final challenge rankings. Subjects scanned by Vendor D were unseen during model training.

For each subject, we have 2D cardiac image slice acquisitions captured at multiple phases in the cardiac cycle (including end-systole and end-diastole). We train on the 2D images independently because we adopt 2D convolution neural networks in our model. Following \cite{baumgartner2017exploration}, we normalize the training data intensity distribution to a Gaussian distribution with a mean of 0 and a standard deviation of 1. Overall, there are 1,738 pairs of images and masks from A and 1,546 pairs of images and masks from B. Apart from these labeled images, there are 47,346 unlabeled images from A, B and C.

\subsection{Model Training}

 Models are trained using the Adam~\cite{kingma2014adam} optimizer with an initial learning rate of 0.001. To stabilize the training, we set the new learning rate to 10\% of the previous rate when the validation Dice similarity coefficient between the predicted and the ground-truth masks does not improve for 2 consecutive epochs. Following the original training setting of SDNet, we set the batch size to 4 and train the model for 50 epochs. All models are implemented in PyTorch~\cite{paszke2019pytorch} and trained using an NVidia 1080 Ti GPU.

\subsection{Results and Discussions}
\label{Sec::Results}

To verify the effectiveness of the proposed augmentation methods, we train 5 models for the purpose of ablation: \textbf{a)} the U-Net model using samples augmented with RA (U-Net+RA), \textbf{b)} the original SDNet model trained in a fully supervised fashion (FS SDNet), \textbf{c)} the fully supervised SDNet model using samples augmented with RA (FS SDNet+RA), \textbf{d)} the SDNet model trained in a semi-supervised fashion, using samples augmented with RA (SS SDNet+RA), and \textbf{e)} the semi-supervised SDNet using samples augmented with both FA and RA (SS SDNet+RA+FA). We train the fully supervised models with labeled samples from vendors A and B, while in the semi-supervised scenario we use all available data (labeled and unlabeled) for training. Table~\ref{tab1} reports the per vendor average Dice scores.

Since we are only allowed to validate our model a limited number of times, the results are not comprehensive, therefore we will do pairwise analysis below.

\begin{table}[t]
\centering
\caption{Evaluation of the 5 models. Average Dice similarity coefficients are reported. Bold numbers denote the best performances across the 5 models. LV: left ventricle, MYO: left ventricular myocardium and RV: right ventricle.}\label{tab1}
\begin{adjustbox}{max width=1\textwidth,center}
\begin{tabular}{|c|c|c|c|c|c|c|c|c|c|c|c|c|}
\hline
\multirow{ 2}{*}{Model} & \multicolumn{3}{c|}{Vendor A} & \multicolumn{3}{c|}{Vendor B} & \multicolumn{3}{c|}{Vendor C} & \multicolumn{3}{c|}{Vendor D} \\ \cline{2-13}
 & LV & MYO & RV & LV & MYO & RV & LV & MYO & RV & LV & MYO & RV\\ 
\hline
U-Net+RA        & 0.900          & 0.829          & 0.811 
                & 0.937          & 0.877          & 0.907 
                & 0.856          & 0.837          & 0.852 
                & 0.762          & 0.651          & 0.503 \\
\hline
FS SDNet        & 0.901          & 0.837          & 0.822 
                & 0.942          & 0.877          & 0.920 
                & 0.851          & 0.826          & \textbf{0.853} 
                & 0.734          & 0.618          & 0.474 \\
\hline
FS SDNet+RA     & 0.905          & 0.846          & 0.828 
                & \textbf{0.945} & 0.886          & \textbf{0.921} 
                & 0.855          & 0.843          & 0.843 
                & \textbf{0.819} & 0.749          & \textbf{0.624} \\
\hline
SS SDNet+RA     & \textbf{0.909} & \textbf{0.854} & \textbf{0.841} 
                & \textbf{0.945} & \textbf{0.887} & 0.916 
                & 0.843          & 0.837          & 0.813 
                & 0.811          & \textbf{0.752} & 0.553 \\
\hline
SS SDNet+RA+FA  & \textbf{0.909} & 0.846          & 0.778 
                & 0.939          & 0.882          & 0.909 
                & \textbf{0.863} & \textbf{0.847} & 0.843 
                & 0.812          & 0.712          & 0.498 \\
\hline
\end{tabular}
\end{adjustbox}
\end{table}

\subsubsection{Does RA help?} By inspection of the \textbf{FS SDNet} and \textbf{FS SDNet+RA} results, we observe that for Vendor A and B, RA helps to achieve better overall performance compared to the respective baseline and RA substantially boosts performance on the unseen Vendor D. Subsequently, for Vendor C, the models have similar performance in the LV class, while FS SDNet+RA achieves the best performance in the MYO class (0.843 Dice score). However, the models using RA do not perform well in the RV class. In the case of Vendor C (no labelled examples), we know that the sample resolution of 24 out of 25 subjects made available at training time from Vendor C is 2.203 mm$^2$, and the resolution of 32 out of 75 Vendor A training subjects is around 2.000 mm$^2$, thus we argue that the resolution bias for Vendor A and Vendor C are already similar in the original data, and becoming invariant to this bias by rescaling the Vendor A data does not further help the performance on Vendor C.

\subsubsection{Does FA help?} Inspection of \textbf{SS SDNet+RA} and \textbf{SS SDNet+RA+FA} results shows that FA has mixed performance. It performs well on Vendor C for the scenario of domain adaptation (target samples available but unlabeled), where Vendor C is one of the vendors providing modality factors. However, FA performs poorly on Vendor D for domain generalization (unseen target), where Vendor D is not involved in the augmentation process.

\subsubsection{Best model:} Out of our two augmentation methods, we observe that RA has the most reliable performance, and we choose to submit the \textbf{FS SDNet+RA} model for final evaluation in the M$\&$Ms challenge. We can see by comparing \textbf{U-Net+RA} with \textbf{FS SDNet+RA} that our chosen architecture of SDNet has good generalization performance compared to a standard U-Net, and this is particularly evident for the unseen vendor D, supporting our choice of the disentangled SDNet architecture even when FA is not employed.

\subsubsection{Improvements to FA:} Satisfyingly, our FA method yields a benefit for Vendor C (whilst RA did not give a significant benefit to Vendor C). However, FA does not give considerable improvements on domain generalization. It may be that improvements to the model training would further enhance the benefit of FA. On the other hand, we should also distinguish the effect of simply training on reconstructed images from the effect of FA. In future, we believe that this method can be extended to achieve better domain generalization once we can manipulate the factors realistically to generate out-of-distribution samples.

\section{Conclusion}
\label{Sec::Conclusion}
In this paper, two data augmentation methods were proposed to address the domain adaptation and generalization problems in the field of CMR image segmentation. In particular, a geometry-related augmentation method was introduced, which aims to remove the scale and resolution bias of the original data. Further, the second proposed augmentation method aims bridge the gap between populations and data captured by different scanners. To achieve this, the original data is projected onto a disentangled latent space and generates new samples by combining disentangled factors from different domains. The presented experimental results showcase the contribution of the geometry-based method to CMR image segmentation through boosting the domain generalization, while also demonstrating the contribution of the disentangled factors mixing method to the domain adaptation. 

\section{Acknowledgement}
The authors of this paper declare that the segmentation method they implemented for participation in the M$\&$Ms challenge has not used any pre-trained models nor additional MRI datasets other than those provided by the organizers.

This work was supported by the University of Edinburgh, the Royal Academy of Engineering and Canon Medical Research Europe by a PhD studentship to Xiao Liu. This work was partially supported by the Alan Turing Institute under the EPSRC grant EP/N510129/1. We thank Nvidia for donating a Titan-X GPU. S.A. Tsaftaris acknowledges the support of the Royal Academy of Engineering and the Research Chairs and Senior Research Fellowships scheme.

\newpage
\bibliographystyle{splncs04}
\bibliography{references}
\end{document}